# Region Based Energy Balanced Inter-cluster communication Protocol for Sensor networks

Rohini Sharma*, D. K. Lobiyal
School of Computer & System Sciences,
Jawaharlal Nehru University
New Delhi, India
rohinisharmaohlan@gmail.com

*Abstract*—Wireless sensor networks faces unbalanced energy consumption problem over time. Clustering provides an energy efficient method to improve lifespan of the sensor network. Cluster head collects data from other nodes and transmits it towards the sink node. Cluster heads which are far-off from the sink, consumes more power in transmission of information towards the sink. We propose Region Based Energy Balanced Inter-cluster communication protocol (RBEBP) to improve lifespan of the sensor network. Monitored area has been divided into regions; cluster heads are selected from specific region based on the residual energy of nodes in that region. If energy of nodes of the specific region is low, nodes from another region are selected as cluster heads. Optimized selection of cluster heads helps in improving lifespan of the sensor network. In our scheme, cluster heads which are far-off from the sink use another cluster heads as the relay nodes to transmit their data to the sink node. So energy of cluster heads deplete in a uniform way and complete area remain covered by sensor nodes. Simulation results demonstrate that RBEBP can effectively reduce total energy depletion and considerably extend lifespan of the network as compared to LEACH protocol. RBEBP also minimize the problem of energy holes in monitored area and improve the throughput of the network

I. INTRODUCTION

Wireless sensor networks or sensor networks are organized of sensor motes which are tiny in size, capable of sensing and communication [1]. Motes (sensor nodes or sensors) are randomly distributed in the area to be monitored and these motes can communicate with each other or with a sink node directly. Sensors sense and transmit the sensed information towards the sink either directly or through multi hop communication. Sensors have limited power batteries and it is not easy to recharge or replace them. So an energy efficient mechanism must be used to improve lifespan of sensor networks. In the multi-hop communication nodes far-off from the sink node uses next node to forward its data, so nodes near the sink diminish their power faster than the nodes far away from the sink node. After some time sensors near the sink exhaust their power and these sensors are not capable to transmit any data. This situation is known as energy holes problem [2]. A detailed survey on energy holes avoiding techniques has been done in [3]. Lifespan of the network can be defined in various ways: it may be defined as the period till the first node expires. Although expiry of a single sensor mote does not influence the observing competencies of the other sensors in the network but it may affect the coverage of one particular area. In the direct communication nodes which are far-off from the sink, may die earlier. During clustering process cluster head can be in any part of the network and can directly move data to the sink, if these cluster heads are far-off from the sink node then they will lose their power faster than other cluster head sensors. So we propose a region balanced protocol which assists balanced energy consumption through the network.

We have used a mix routing approach where a cluster head can directly transmit data to the sink node or it may use other cluster heads to transmit its data to the sink. The other cluster heads act as relay node. The decision depends on the distance of cluster head from the sink. If the cluster head is close to the sink it can communicate directly otherwise it take help of other cluster heads. So we have used an approach in which distance between cluster head and sink is reduced by using other cluster heads. Here another problem known as energy holes may arises as cluster head nodes which are chosen as relay nodes and are nearer to the sink node depletes their energy quicker. To avoid this situation we have divided the region in to inner and outer circular regions. Cluster heads which work as the relay nodes are chosen based on their position in the area and remaining energy. So lifespan may also be defined as the period till a ratio of nodes expires. If number of nodes expiry increases above a certain limit some regions of the area can be uncovered or network may be partitioned. The proposed method assists in the uniform consumption of energy through the network. The proposed method has been compared with famous LEACH protocol [4] and it is found that proposed method has momentous enhancement over LEACH in provisions of lifespan, energy consumption rate and throughput of the network for different node density. The rest of the paper is systematized as follows. Section 2 gives an outline of correlated work in the literature. Section 3 describes network deployment model. Section 4 explains energy consumption model. Section 5 explains unbalanced energy consumption issues and the proposed solution. Section 6 includes simulation environment, results and their analysis. Section 7 concludes the work.

## II. RELATED WORK

There are two types of routing protocols, flat [5] and hierarchical protocols [6-8]. Hierarchical routing protocols are better than flat routing protocols as they offer an energy balanced routing and scalability to sensor networks. There are various cluster based energy efficient protocols available in the literature. Chao et al. have proposed clustering based method in which member nodes switched in to sleep state if the currently sensed data is same as the previously sensed data. It helps in improving transmission energy efficiently [9]. Wang et al. have projected a multi-hop intra cluster routing protocol by using conditions of residual energy and distance [10]. Rani et al. have proposed EEICCP protocol which used multi-hop approach for CHs (cluster heads) to transmit data to the sink [11]. Authors in [12] have proposed an energy aware intra cluster algorithm for sensor networks. Authors in [13] have selected a set of cluster heads for forthcoming rounds based on the average energy of cluster and residual power of nodes. Authors in [14] have proposed clustering threshold for normal and advanced nodes in heterogeneous sensor network to improve network stability and lifetime. None of the above methods tried to minimize distance between cluster heads and base station. Sinha et al. have proposed entropy based clustering method in which motes sensing the same type of data are positioned in separate cluster [15]. Singh et al. have proposed load balancing clustering method for energy efficiency in sensor networks [16]. Authors in [17] have used sleep scheduling policy to balance energy consumption with in a cluster. When distance between the sink and a cluster head increase, cluster head consumes more energy to transmit the data. To solve this issue Multi-hop LEACH is proposed in [18]. But they have not considered the issue of overused relay cluster heads. Aslam et al. have given a detailed survey on LEACH based routing protocols [19]. Our method proposes a region based energy efficient inter cluster communication method which improve lifetime and throughput of the network. Fig. 1 explains the working of the proposed model.

## III. NETWORK DEPLOYMENT MODEL

For designing sensor network, following assumptions are considered:
- All sensors are homogeneous in provisions of initial energy.
- All sensor nodes are stationary after deployment.
- Nodes are randomly distributed in monitored area.
- Nodes are location aware and are capable of transmission data directly to the sink.
- Receiver sensor can determine distance from sender by using received signal strength indicator.
- The sink node is in outer area of the network zone

## IV. ENERGY CONSUMPTION MODEL

We have taken the first order radio energy model [4] to determine energy consumption among motes. More energy is consumed in transmission than in receiving. If distance between sender and receiver is $d$ and total $k$ bits are being transmitted then total consumption of energy in transmission is given by:

$$E_{TX} = \begin{cases} k*E_{elec} + k*\varepsilon_{fs-amp}*d^2, d<d_o \\ k*E_{elec} + k*\varepsilon_{tworay-amp}*d^4, d \geq d_o \end{cases} \quad (1)$$

Here $d_0$ is a cross over point (threshold value) and $E_{elec}$ is the energy consumed by a node in active mode [4]. If distance $d$ between two nodes is less than $d_0$ free space model is used for energy consumption otherwise two ray multi path fading channel is used for energy consumption. Energy consumed in receiving $k$ bits of data is given by:

$$E_{RX}(k) = E_{elec} \times k \quad (2)$$

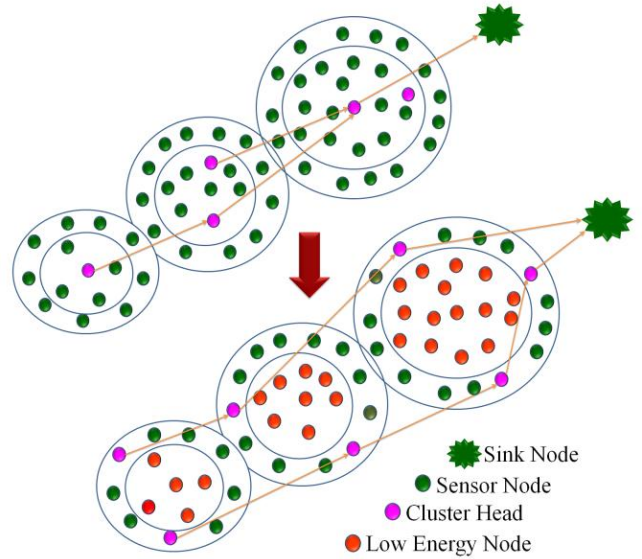

Fig. 1. Division of area into circular regions and selection of Cluster heads.

## V. ENERGY UNBALANCED CONSUMPTION ISSUES AND PROPOSED SOLUTION

### A. Energy Unbalanced Consumption Problem

In wireless sensor networks battery power is limited. So energy must be consumed in such a way that network can work for a long time. In a multi-hop communication, sensors adjacent the sink node utilize more power while in a direct hop communication, nodes which are far-off from the sink consume more power. A multi-hop communication may leads to energy holes near the sink and a direct communication may lead to energy holes in outer region of the network. If energy holes are created near the sink node, no other data can be transmitted to sink node and remaining energy of the network will be wasted. If energy holes are created in outer region then outer area of network will remain uncovered. Our method provides uniform and balanced energy consumption in the network.

## B. Proposed Model

First of all divide the region in two circular inner and outer regions. Keep inner region larger than outer region as shown in Fig.1. Randomly distribute the nodes in the area; more nodes will be in inner region than the outer region because size of inner region is larger than outer region. Area between two concentric regions is called annulus. It can be calculated as follows:

$$Area = \pi(R^2 - r^2) \qquad (3)$$

Where $R$ is the radius of outer circle and $r$ is the radius of inner circle. If the value of $R$ and $r$ differ at most by 1, then the value of annulus will be very small. Greater the difference between $R$ and $r$, smaller will be the inner region.

Now distributed clustering process starts and cluster heads are selected to gather data from the other sensor nodes. Initially cluster heads are selected from the inner regions of the network, these CHs send advertisement message to neighboring nodes, sensor nodes send response message and data is transferred according to TDMA scheduling [4]. Member nodes can be from inner region or outer region area. Now steady phase begins and CHs start transmitting aggregated data, CHs which are closest to sink node, transmit data straight to the sink but CHs which are far-off from the sink, transmit data to subsequent nearest CH in the upper region. Let $S$ is the sink node. In data transmission form cluster head $CH_i$ to cluster head $CH_j$ the energy consumption is given by:

$$E_{TX}(CH_i, CH_j) = \\ E_{TX}(k, d(CH_i, CH_j)) + E_{RX}(k) + E_{TX}(k, d(CH_j, S)) \qquad (4)$$

Cluster head with smallest value of $E_{TX}$ will be selected as relay cluster head. Its value primarily depends on distance between communicating cluster heads and sink node. Further to avoid energy holes nearby sink node, we have considered remaining energy of relay cluster head. If the remaining energy of relay $CH_j$ is smaller than $CH_k$ and distance of both cluster heads towards the sink is same then we chose $CH_k$ as the relay cluster head. The complete process of inter cluster communication has been shown in Fig 2.

After every round residual or left over energy of the nodes is calculated. If the total energy of nodes in the inner circular region is lesser than the overall energy of nodes in the outer region, CHs are selected from the outer circular region. For initial rounds total energy of nodes in inner region will be more than total energy of nodes in outer region because density of nodes is more in inner region as compared to outer region. Inner region is considerably large and initially CHs are chosen from this area. Cluster head selection plays a vital function in energy efficacy of the network. Nodes in this region are at an optimal distance from the sink and from nodes in surrounding area. Further CH nodes from this area act as relay nodes; CHs carries more data and consumes more energy, energy holes can emerge in this area. To avoid energy holes we have a set of nodes which can act as CH in upcoming rounds. These set of nodes are placed with in inner area and these nodes also have optimal or near optimal distance towards the sink and other nodes in the area. So set of CHs and candidate CHs is large enough for some rounds. CHs in the inner region deplete less energy because they are at optimal distance and number of alternate CHs are much larger.

As shown in Fig. 1, when energy of nodes in inner region becomes lower than energy of nodes in outer region, CHs from outer regions are selected. Nodes in the inner region have lesser energy than nodes in outer region but still nodes in inner region can sense and transmit data to newly selected head. As these nodes are near to CHs, transmission distance of these nodes is less. Data transmission towards the sink is the function of CHs. In this way area will remain covered by sensor nodes and energy holes do not emerge in the area. As shown in Fig. 2 CH nodes (red color) of outer most area send their data through CH nodes in the region nearer to the sink. Initially three nodes from inner regions are selected as CHs, 1st CH transmits data to $2^{nd}$ CH which in turn transmits data to 3rd CH and finally last CH transmits data to the sink or base station. In next round another mote with maximum residual power is chosen as CH in the inner circular region until total energy in the inner circular region becomes less than total energy in the outer circular region.

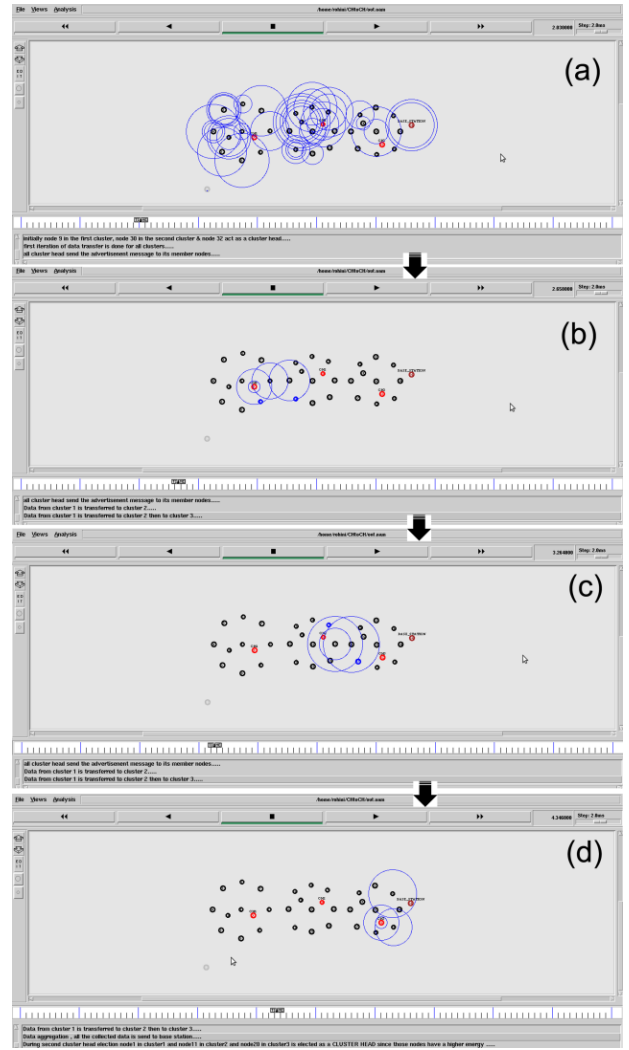

Fig. 2. Simulation view of inter-cluster communication.

## C. Flow Chart of the Proposed Solution

Fig. 3 explains the complete process of RBEB protocol through the flow chart.

## VI. RESULTS AND ANALYSIS

### A. Simulation Environment

Proposed solution has been designed and simulated in ns-2.34 [20-21]. RBEBP has been compared with LEACH [4] protocol in terms of network lifespan, energy consumed and total data units received (throughput) at the sink node. Simulation parameter and their values have been given in TABLE I. Protocol performance is measured for different node density.

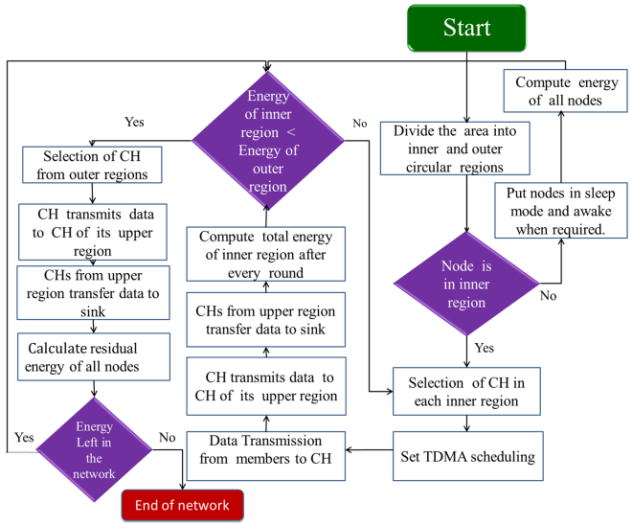

Fig. 3. Flowchart for proposed solution.

TABLE I. SIMULATION SETUP PARAMETERS

| S.No. | Parameters | Values |
|---|---|---|
| 1. | Area | 1000 m × 1000 m |
| 2. | Number of Nodes | 35, 50, 100 |
| 3. | Node Mobility | No |
| 4. | Traffic | CBR (bits/sec) |
| 5. | Initial energy of nodes | 2 Joules |
| 6. | Sink node position | (75, 175) |
| 7. | $\varepsilon_{fs}$ (free space model energy consumption) | 10 pJ/bit/m$^2$ |
| 8. | $\varepsilon_{mp}$ (multi path model energy consumption) | 0.0013 pJ/bit/m$^4$ |
| 9. | Cross over point $d_o$ | $(\varepsilon_{fs}/\varepsilon_{mp})^{1/2}$ m |

### B. Simulation Results analysis

Performance of RBEB is measured against different performance metrics as shown below:

*1) Network Lifespan*: Lifespan or lifetime of the network has been defined as the time between the start and end of the network operation. It is measured by the number of nodes alive in the network. It can be further categorized by the number of nodes alive in the network. Fig. 4 gives the number of nodes alive per sec in the network for both protocols. Fig. 5 gives the energy consumption behavior of protocols.

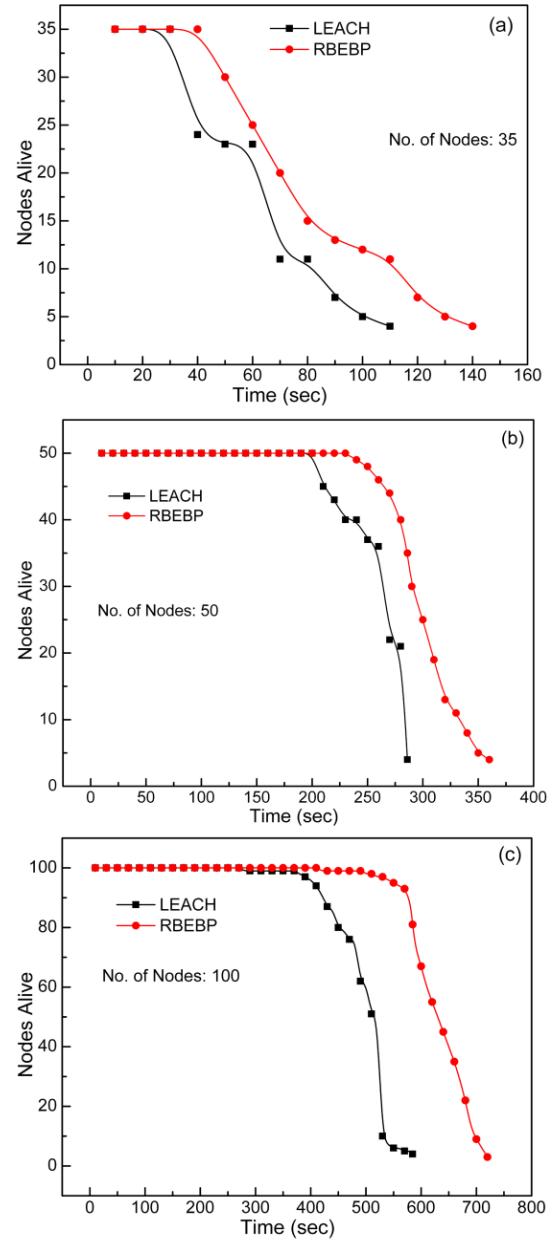

Fig. 4. Number of nodes alive in the network per sec for different node density.

First Node Die (FND): This is time gap between the commencement of network process and death of first node. It is also known as stability period. After first node death, area covered by that node may remain uncovered if node density is low, however if node density is very high that area can be covered by some other node.

Half Nodes Die (HND): It is the time gap between the commencement of network process and death of 50% of the nodes of the network. It is an important parameter as now less hops are available for communication up to sink node, less area will be covered and network performance effects severely.

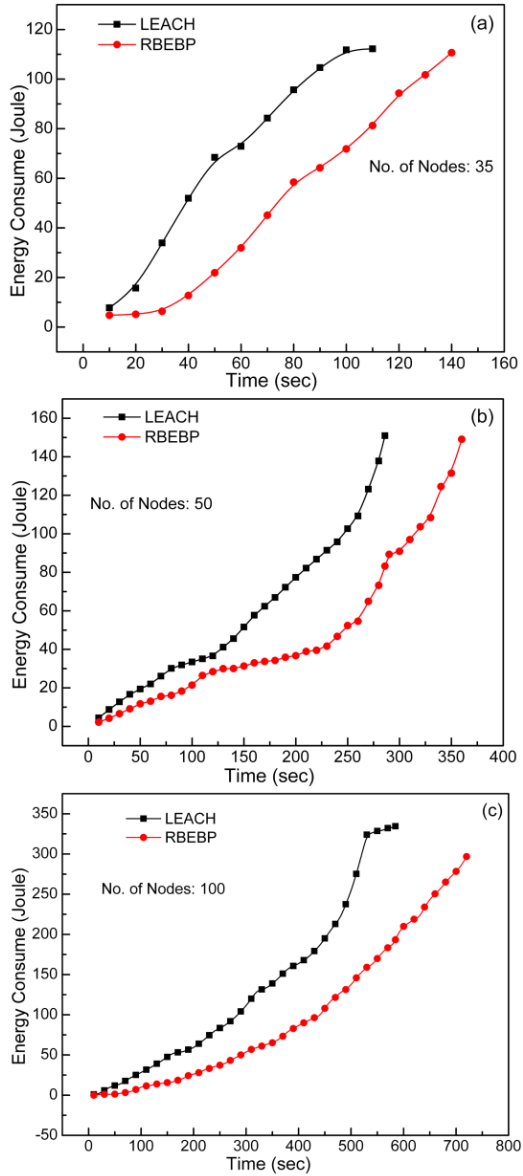

Fig. 5. Energy consumption of network for RBEBP and LEACH protocol.

All Nodes Die (AND): It the time gap between the commencement of the network process and death of last node. It concludes that no node is alive in the network and network cannot function any more.

A protocol performance is considered good if it works for a longer time and consumed the available energy in the network efficiently. It is evident from Fig. 4 that network works for longer time in presence of RBEBP method. As nodes density increase operation time of both the protocols increases but RBEBP outperforms LEACH protocol. It runs for longer time as compared to LEACH protocol. For 35 nodes, it runs for 140 sec, for 50 nodes it run for 360 sec and for 100 nodes it runs for 720 sec. TABLE II illustrates network lifespan in terms of AND, FND and HND metrics. It is clear from TABLE II that duration and number of live nodes is more for RBEBP as compared to LEACH protocol. Fig. 6 gives total throughput of network.

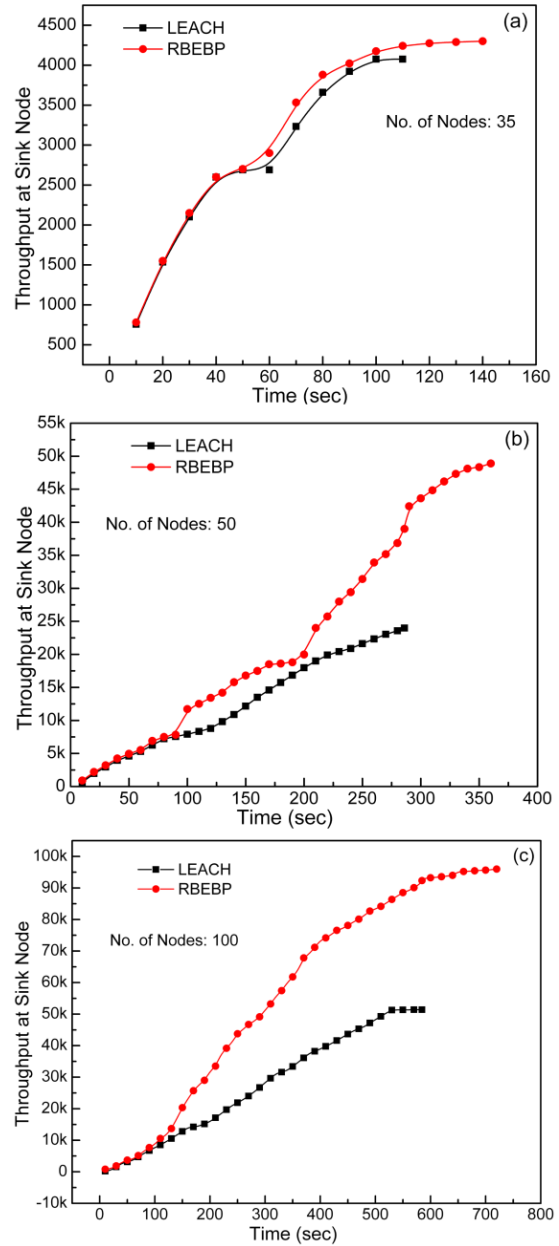

Fig. 6. Throughput of network for RBEBP and LEACH protocol.

*2) Energy Consumed in the network*: It the quantity of energy consumed per sec in the network. It is illustrated from Fig. 5 that amount of energy consumed is lesser for RBEB as compared to LEACH protocol.

RBEBP consumes less energy and maintain the operation of network for longer time as compared to LEACH. For higher number of nodes initially amount of energy consumed is same but with time energy consumption in LEACH becomes more as compared to RBEBP

*3) Throughput at the sink node*: It is the quantity of total data arriving at the sink node. It is described from Fig. 6 that amount of data arriving at the sink is more for RBEBP as compared to LEACH. As RBEBP consumes energy in a uniform and balanced way, more nodes are alive in the network for a longtime and therefore these node produce throughput of the network. Although amount of throughput does not differ much for small node density for both protocols but RBEBP produces throughput for long time.

TABLE II. NETWORK LIFESPAN COMPARISON FOR DIFFERENT NODES DENSITY

| Protocol | No. of Nodes | FND (sec) | HND (sec) | AND (sec) |
|---|---|---|---|---|
| LEACH | 35 | 40 | 65 | 110 |
| | 50 | 205 | 262 | 290 |
| | 100 | 280 | 520 | 590 |
| RBEBP | 35 | 50 | 75 | 140 |
| | 50 | 240 | 30 | 364 |
| | 100 | 420 | 630 | 720 |

## VII. CONCLUSIONS

Clustering is an efficient method to improve lifespan of sensor networks. Optimal cluster heads selection improves the efficiency of basic clustering algorithm. RBEBP select those nodes as a cluster heads which are at an optimal distance from the sink node. This position of cluster heads consumes less energy in transmission. Further cluster heads chose another cluster heads to transmit their data; it reduces communication distance between sender and receiver. RBEBP can make a balance between distance and residual energy during the set up phase of clustering. Simulation results validates that RBEB can successfully minimize energy consumption and elude the creation of energy holes in the network. It improves network lifespan and throughput at the sink node. For RBEBP time of first node, half nodes and all nodes death is more as compared to LEACH protocol. It means network have alive nodes for a long time.

ACKNOWLEDGEMENT

Rohini Sharma is thankful to the Council of Scientific and Industrial Research (CSIR), Human Resource Development Group, India for providing necessary fellowship.